\documentclass[prl,twocolumn,showpacs]{revtex4}

\usepackage{graphicx}
\usepackage{amsmath,amssymb}
\usepackage{amsbsy}
\usepackage{bm}

\begin{document}

%%%%%%%%%%%%%%%%%%%%%%%%%%%%%%%%%%%%%%%%%%%%%%%%%%%%%%%%%%%%%%%%%%%%%
%%%%%%%%% new commands %%%%%%%%%%%%%%%%%%%%%%%%%%%%%%%%%%%%%%%%%%%%%%
%%%%%%%%%%%%%%%%%%%%%%%%%%%%%%%%%%%%%%%%%%%%%%%%%%%%%%%%%%%%%%%%%%%%%

\renewcommand{\d}{{\rm d}}
\newcommand{\e}{{\rm e}}
\newcommand{\imai}{{\rm i}}

%%%%%%%%%%%%%%%%%%%%%%%%%%%%%%%%%%%%%%%%%%%%%%%%%%%%%%%%%%%%%%%%%%%%%%%
%%%%%%%%%%%%%%%%%%%%%%%%%%% Abstract %%%%%%%%%%%%%%%%%%%%%%%%%%%%%%%%%%
%%%%%%%%%%%%%%%%%%%%%%%%%%%%%%%%%%%%%%%%%%%%%%%%%%%%%%%%%%%%%%%%%%%%%%%

\title{Zero-phonon linewidth and phonon satellites in the
optical absorption of nanowire-based quantum dots}

\author{Greta Lindwall}\altaffiliation[Now at ]{Corrosion and Metals 
Research Institute, Drottning Kristinas v{\"a}g 48, 114 28 Stockholm, Sweden}
\author{Andreas Wacker}
\email{Andreas.Wacker@fysik.lu.se}
\affiliation{Mathematical Physics, Lund University, Box 118,
22100 Lund, Sweden}
\author{Carsten Weber}
\author{Andreas Knorr}
\affiliation{Institut f\"ur Theoretische Physik, Technische Universit{\"a}t
Berlin, Hardenbergstr. 36, 10623 Berlin, Germany}
\date{30. July 2007, to appear in Physical Review Letters}

\begin{abstract}
The optical properties of quantum dots embedded in a
catalytically grown semiconductor nanowire
are studied theoretically. In comparison to dots in a bulk environment, 
the excitonic absorption is strongly modified by the
one-dimensional character of the nanowire phonon spectrum.
In addition to pronounced satellite peaks due to phonon-assisted absorption, 
we find a finite width of the zero-phonon line already in the lowest-order
calculation.
\end{abstract}

\maketitle

Decoherence in semiconductor heterostructures such as quantum dots are
of central importance due to their use in advanced photonic structures
such as emitters of single and entangled photons
\cite{MoreauPRL2001,KakoNatureMat2006} for quantum cryptography and
in photonic crystals for solid state cavity QED
\cite{HennessyNature2007}. In particular, the stochastic modulation of
the phase by a bath via electron-phonon scattering constitutes an
important source of optical decoherence. In this Letter, we show that
the dimensionality of the semiconductor structure embedding
the quantum dot essentially affects the dephasing properties.

Catalytically grown semiconductor nanowires \cite{AgarwalAP2006} allow
for the inclusion of quantum dots which exhibit a zero-dimensional
density of states as probed by conductance measurements
\cite{FranceschiAPL03,ThelanderAPL03}. Here, we focus on the optical
absorption line-shape of the transition between the highest
valence-band state and the lowest conduction-band state within the
quantum dot (the lowest quantum confined interband transition in the
dot), where a clear signal from the bound exciton can be observed in
the photoluminescence spectrum \cite{PanevAPL2003}. Intensive
single-photon emission was demonstrated suggesting potential
applications in quantum information processing of nanowire-based
quantum dots \cite{BorgstromNL2005}.

The optical absorption of electromagnetic radiation by this two-level
transition is mediated via a coherent superposition of the initial
and final state, the polarization. The temporal coherence decay of
the polarization (dephasing time) is directly reflected in the
line-shape of the optical absorption as a function of frequency. For
quantum dots, the discreteness of energy levels does not allow for
standard scattering transitions known from bulk material, where the
phonon energy matches the energy difference between initial and final
states. Thus, long dephasing times of several hundred picoseconds
have been observed \cite{BorriPRL2001} in a sample of self-organized
quantum dots \cite{Bimberg1999}. In addition, calculations for such
structures show that scattering of electrons with phonons of the
embedding material leads to sidebands in the excitonic spectra which
are due to phonon-assisted processes occurring on a shorter time scale
\cite{KrummheuerPRB2002,ForstnerPRL2003,ForstnerPSSB2003,Villas-boasPRL2005,KrugelPRB2006},
features which have also been observed experimentally
\cite{BesombesPRB2001,BorriPRL2001,FaveroPRB2003}. The width of the
central absorption peak itself, the zero-phonon line, is an issue of
particular interest, as standard approaches such as the independent
Boson model
\cite{KrummheuerPRB2002,MahanBook2000,ZimmermannICPS2002,ForstnerPSSB2002}
provide a zero width in contrast to the experimental findings
\cite{BorriPRL2001,BorriPRB2005}. These works focused on quantum dots
embedded in a bulk system as appropriate for the self-organized dots
addressed in the experiment \cite{BorriPRL2001}. 
In addition, etched mesas with small
diameters down to the 200 nm range have been studied, where the change
in zero-phonon linewidth was attributed to phonon scattering at the
etched walls \cite{OrtnerPRB2004,RudinPRB2006}.

In contrast, we are focusing here on dots embedded in catalytically grown
nanowires with a diameter in the 50 nm range. These wires
have a very regular structure \cite{BjorkNano2002},
and thus we expect significantly longer lifetimes for the 
quasi one-dimensional phonon modes. We show that the dimensionality
of the phonon spectrum essentially modifies the optical absorption 
of the embedded quantum dot. In particular, for nanowire-based quantum
dots, a finite width of the zero-phonon line is found within the 
independent Boson model.

\begin{figure}
 \includegraphics[width=8.4cm]{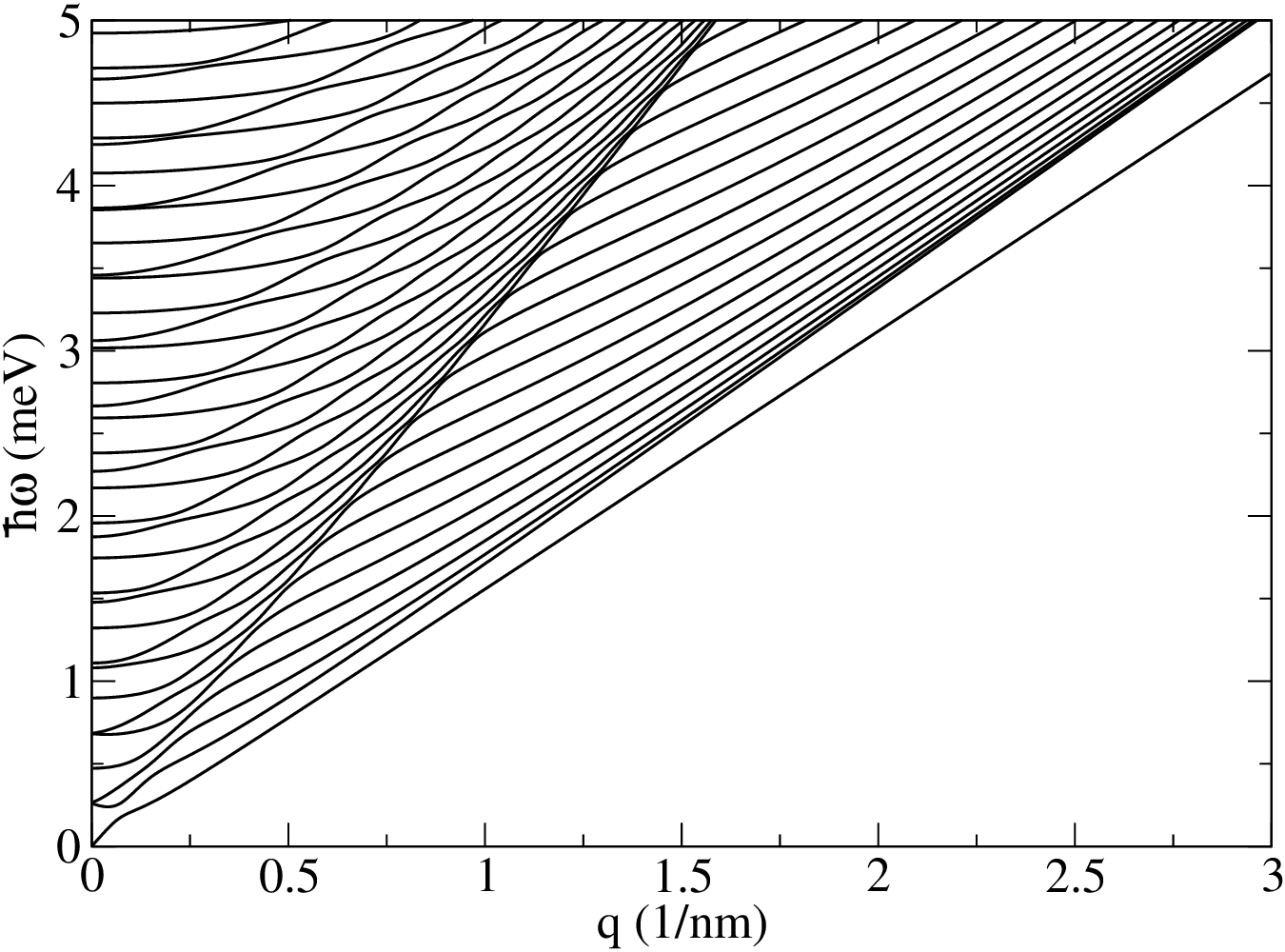} 
 \caption{Dispersion relation $\hbar\omega_{q\kappa}$
  for the acoustic phonons in a
  cylindrical GaAs nanowire with a radius of 25 nm. (Parameters:
mass density $\rho=5370$ kg$/$m$^3$, longitudinal sound velocity
$v_L=4780$ m$/$s, transverse sound velocity $v_T=2560$ m$/$s). Only
the compressional modes
  are shown.}
\label{FigDispersion}
\end{figure}

We approximate the hexagonal GaAs nanowires \cite{LarssonNT2007}
to be cylindrical with a radius of 25 nm and evaluate the phonon
spectrum for the nanowires within an isotropic continuum model
following Ref.~\onlinecite{StroscioJAP1994}. 
This geometry results in a
classification of the phonon modes by a one-dimensional Bloch vector
$q$ and a multitude of modes numbered by $\kappa$ with 
increasing energy. The resulting dispersion $\omega_{q\kappa}$ is
shown in Fig.~\ref{FigDispersion}, where we restricted to the
compressional modes, as only those couple via the 
electron-phonon deformation potential 
interaction to the radially symmetric electron states considered below.

\begin{figure}
 \includegraphics[width=8.4cm]{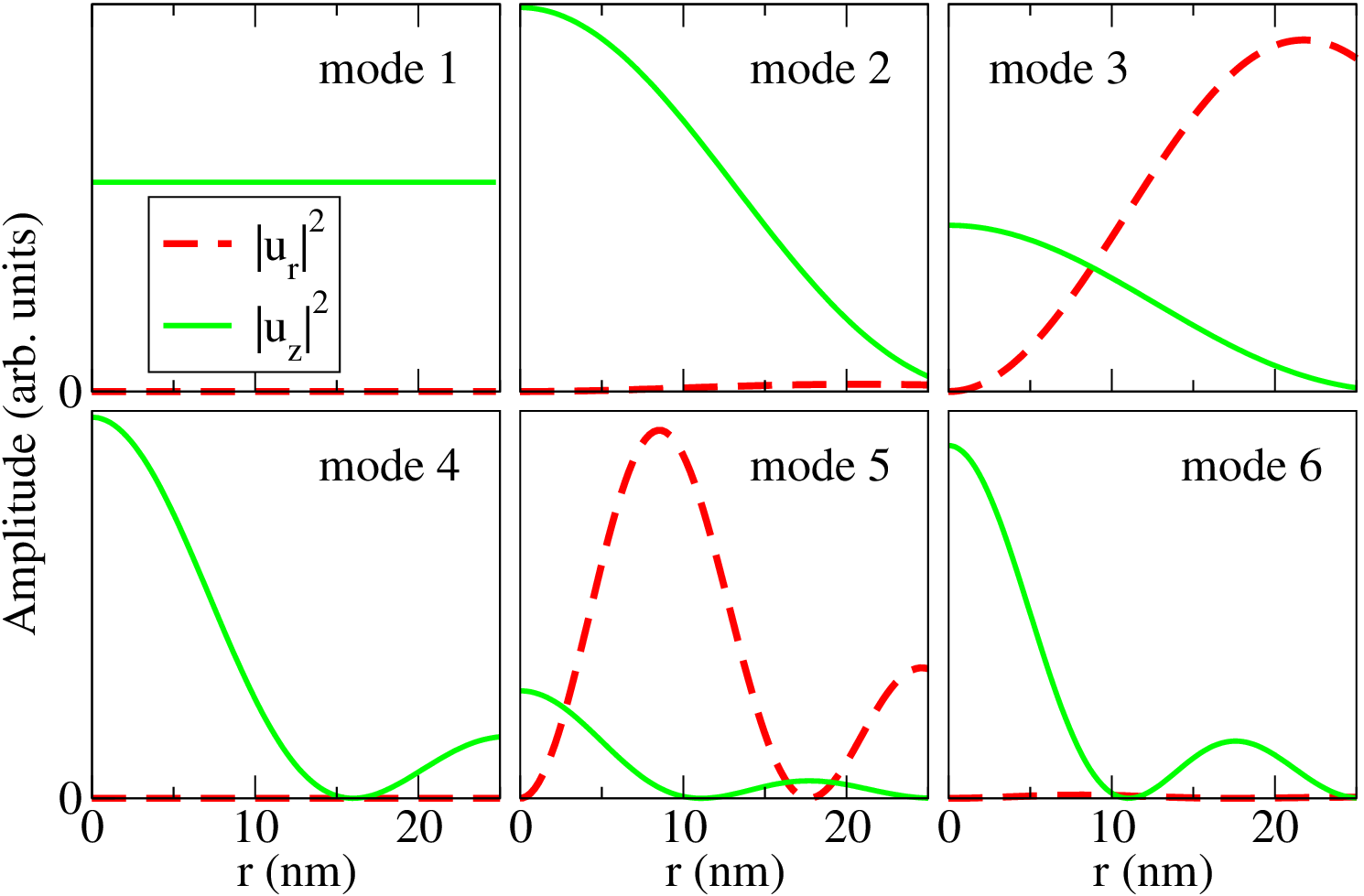} 
 \caption{[color online] Radial $(u_r)$ and axial $(u_z)$ amplitudes 
  as a function
  of the radial coordinate $r$ for the first
  six modes of the dispersion relation from
  Fig.~\ref{FigDispersion}. The amplitudes are displayed for $q=0.003/$nm.}
\label{FigModes}
\end{figure}

Several modes of the phonon spectrum are shown in Fig.~\ref{FigModes}.
Modes 1, 2, 4, and 6 are essentially axial modes, 
where the elongation is parallel
to the wire, while modes 3 and 5 exhibit a significant radial part 
as well. For a
finite wave vector both types mix, which is of particular relevance
for modes 2 and 3, as well as 5 and 6, which are degenerate at $q=0$.

We calculate the absorption from the independent Boson model
following
Refs.~\onlinecite{KrummheuerPRB2002,ForstnerPSSB2002,ZimmermannICPS2002,MahanBook2000,MuljarovPRL2004}. For
a $\delta$-shaped laser pulse applied at $t=0$, the polarization
$p(t)$ between the electron and hole evolves as
\begin{multline}
 p(t) = p(0)\e^{-\imai\omega_{\mathrm{ren}}t}\\
 \times\exp\left[
 -\sum_{q\kappa}\frac{|g_{x}^{q\kappa}|^{2}}{\hbar^{2}\omega_{q\kappa}^{2}}
 \left(4 n_{q\kappa}\sin^2\frac{\omega_{q\kappa}t}{2}
 +1-e^{-i\omega_{q\kappa}t}\right)
 \right]
 \label{EqpolarizationExact}
\end{multline}
for $t>0$,
where $n_{q\kappa}$ is the phonon distribution given by the Bose-Einstein
distribution $n_{q\kappa}=(e^{\hbar \omega_{q\kappa}/k_{B}T}-1)^{-1}$.
The coupling to the phonons renormalizes the transition frequency to
\begin{equation}
\hbar\omega_{\mathrm{ren}}=E_{e}-E_{h}-E_{\mathrm{Coulomb}}-\sum_{q\kappa}
\frac{|g_{x}^{q\kappa}|^{2}}{\hbar\omega_{q\kappa}}\, ,
\end{equation}
where $E_{e}$ and $E_{h}$ are the single-particle electron and hole ground
state energy, respectively, and $E_{\mathrm{Coulomb}}$ is the exciton
binding energy. $\hbar\omega_{\mathrm{ren}}$ is used
as the reference point for the frequency detuning in the following 
absorption spectra.
The coupling to the phonons is reflected by the coupling elements
\begin{equation}
g_{x}^{q\kappa}=\int \d^3 r
\left[D_c|\Psi_e({\bf r})|^2-D_v|\Psi_h({\bf r})|^2\right]
\nabla\cdot {\bf u}_{q\kappa}({\bf r})
\label{Eqg}
\end{equation}
which is the difference between the coupling
of the electron and the hole to the respective phonon mode within
the deformation potential interaction. We use the interaction
parameters $D_c=-14.6\mathrm{eV}$ and $D_v=-4.8\mathrm{eV}$ for the
conduction and valence band, respectively. ${\bf u}_{q\kappa}({\bf
r})$ is the normalized elongation field of the phonon mode. The wave
functions for the electron (e) and hole (h) state are modeled by
spherical Gaussian wave functions $\Psi_\alpha({\bf r})\propto
\e^{-(r^2+z^2)/2a_{\alpha}^2}$ with $a_{e}=5.8$ nm and $a_{h}=3.19$ nm
like in Ref.~\cite{ForstnerPRL2003}. As
the polarization, Eq. (\ref{EqpolarizationExact}), is the response to a
$\delta$-signal, its Fourier transformation provides the complex
susceptibility $\chi(\omega)$ which yields the absorption
spectrum via its imaginary part \cite{HaugKochBook2004}.

The resulting absorption spectrum is shown in
Fig.~\ref{FigAbsorption} both for bulk and wire phonons
using identical parameters.
For bulk phonons, we find a sharp zero-phonon line, which is here
artifically broadened by introducing a finite decay time. This can be
motivated by spontaneous recombination
or temperature-dependent broadening mechanisms discussed in the
literature recently for
acoustic \cite{MuljarovPRL2004,MuljarovPRL2005,MachnikowskiPRL2006}
or optical \cite{MachnikowskiPRB2005, MuljarovPRL2007}
phonons. In addition to the zero-phonon line, rather flat
absorption features (sidebands) are present for detunings
$\hbar|\omega-\omega_{\mathrm{ren}}|\lesssim 1$ meV.
In contrast, for wire phonons, we find distinct satellite peaks in the
sidebands (i) and a finite width of the zero-phonon line without any
artificial broadening (ii), 
which constitute the two central results of this article:

\begin{figure}
\includegraphics[width=8.4cm]{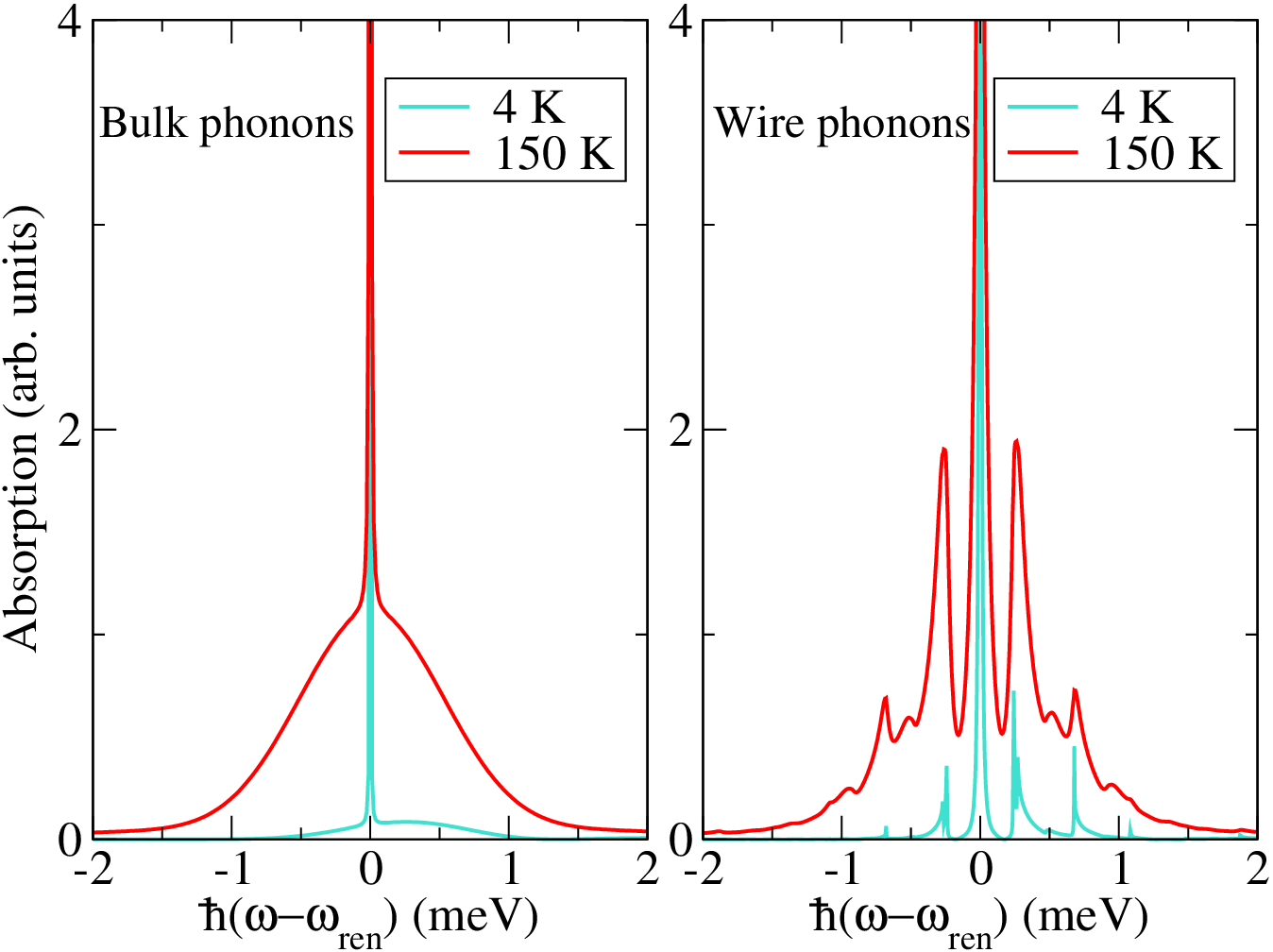} 
 \caption{[color online] Absorption versus detuning from the renormalized
  transition energy for a spherical quantum dot for different
  temperatures. Left: bulk
  phonons (as in Ref.~\onlinecite{ForstnerPRL2003}),
  right: wire phonons from Fig.~\ref{FigDispersion}.}
\label{FigAbsorption}
\end{figure}

\begin{figure}
 \includegraphics[width=8cm]{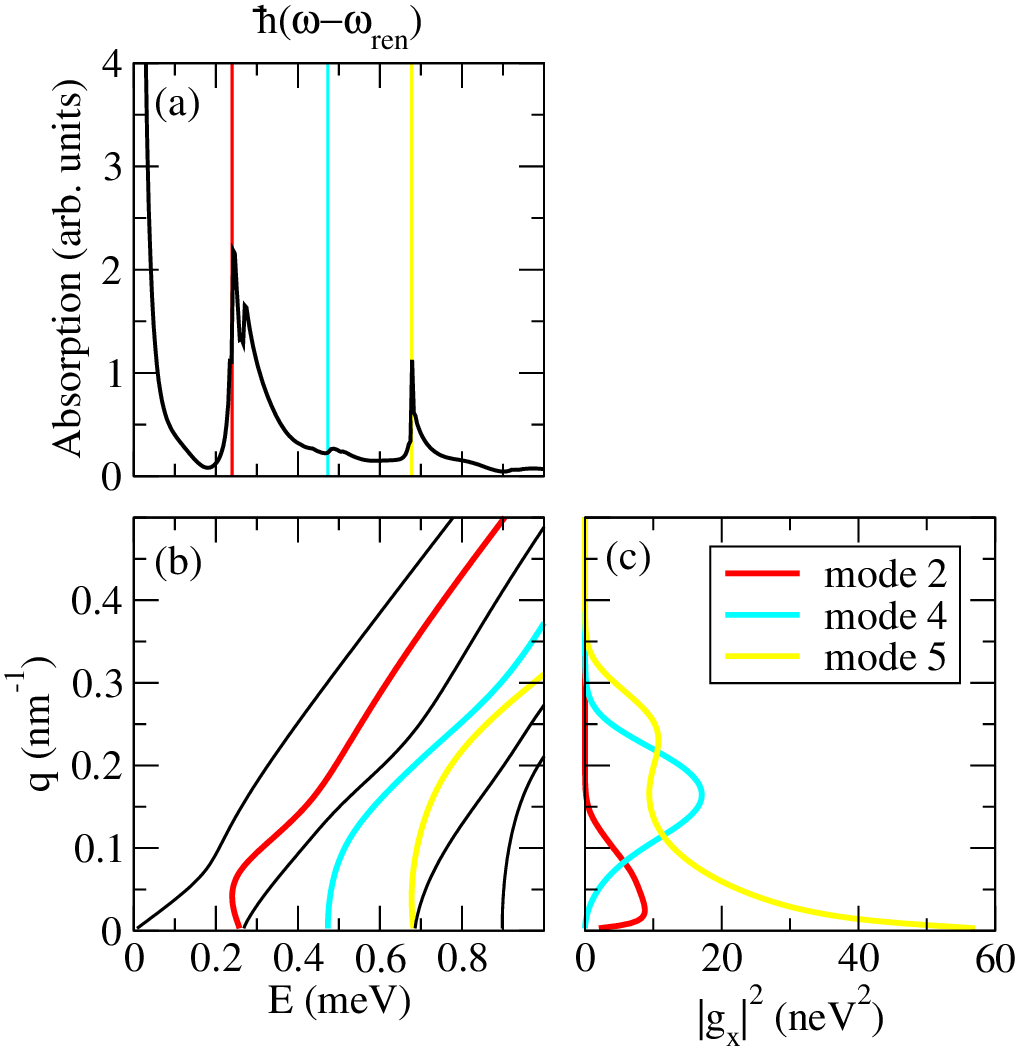} 
  \caption{[color online] (a) Part of the absorption spectrum 
   of the quantum dot embedded in a wire at 40 K. (b) Details of the
   phonon spectrum from Fig.~\ref{FigDispersion} at relevant
   energies. (c) Magnitude of the exciton coupling
   for the second (red), fourth (cyan), and fifth (yellow) mode.}
\label{FigAbsorptionExplain}
\end{figure}

(i) Let us first focus on the satellite peaks,
which are similar to the result of calculations
for a slab-geometry \cite{KrummheuerPRB2005}.
Fig.~\ref{FigAbsorptionExplain} shows that the peaks in the
absorption are related to the extrema in the phonon dispersion. At
these extrema, the phonon density of states has singularities.
However, not all phonon modes contribute, as their coupling elements
differ. E.g., the fourth mode is of axial character,
i.e. ${\bf u}_{q\kappa}({\bf r})\sim u_0(r) {\bf e}_z \e^{\imai qz} $ in
cylindrical coordinates $(r,z)$ with the radial and longitudinal
coordinates of the wire $r$ and $z$, respectively, see Fig.~\ref{FigModes}.
Thus the divergence vanishes with $q$ for
$q\to 0$, and so does the corresponding exciton coupling matrix element
$g_{x}^{q\kappa}$, as shown in Fig.~\ref{FigAbsorptionExplain}(c).
Thus, this mode couples only weakly to the exciton wave function 
around its extremum at $q=0$, and subsequently only a weak peak is 
observed in the
absorption. In contrast, the fifth mode is of radial character leading to
a finite exciton coupling matrix element for $q\rightarrow 0$ and
a strong peak in the absorption.
The second phonon mode is of particular interest, as it
is of axial nature for $q\to 0$, where the coupling vanishes. However,
due to mixing with mode 3 at finite $q$,
its frequency decreases with $q$ for small
$q$, and the minimum of frequency occurs at finite $q$ where its
density of states exhibits a singularity.
Here, the exciton coupling matrix element
is finite so that a strong peak is visible in the absorption.

(ii) Now we turn to the broadening of the zero-phonon line:
The first phonon mode $(\kappa=1)$ is the only mode 
which extends to
$\omega\to 0$, where it describes a wave of elongations
in $z$ direction traveling along the wire:
\begin{equation}
{\bf u}_{q1}({\bf r})
\stackrel{q\to 0}{\sim} N_q \left[\left(1-\nu \frac{q^2r^2}{2}\right){\bf e}_z
-\imai\nu qr{\bf e}_r+{\cal O}(q^3)\right]\e^{\imai qz}\, ,
\end{equation}
where $\nu$ is the Poisson number, which is
the ratio between transverse contraction and longitudinal
elongation in the direction of the stretching force. 
In the limit $q\to 0$, it has a 
linear dispersion $\omega_{q1}=v_1q$ with
$v_1=\sqrt{E/\rho}$, where $E$ is the Young modulus and
$\rho$ the mass density \cite{LandauContMechChap25}.
The proper normalization \cite{MahanBook2000} leads to
\begin{equation}
N_q=\sqrt{\frac{\hbar}{2LA\rho \omega_{q1}}}[1+{\cal O}(q^2R^2)]
\end{equation}
with the wire area $A$ and the normalization length $L$ for the 
phonon modes.
Eq.~(\ref{Eqg}) yields
\begin{equation}
g_{x}^{q1}\sim \imai q\sqrt{\frac{\hbar}{2LA\rho \omega_{q1}}}
(D_c-D_v)(1-2\nu)\quad \mbox{for}\, q\to 0\, .
\end{equation}
Furthermore, in thermal equilibrium $n_{q1}\sim
k_BT/\hbar\omega_{q1}$ holds for $\omega_{q1}\to 0$. For long times, we
evaluate the contribution to the dephasing in the exponential in 
Eq.~(\ref{EqpolarizationExact}) analogously to Fermi's golden rule:
\[
\frac{4\sin^2\frac{\omega_{q\kappa}t}{2}}{\omega^2_{q\kappa}}
\sim 2\pi t\delta(\omega_{q\kappa})\, .
\]
Together, this results in
\begin{equation}
\sum_q\frac{|g_{x}^{q1}|^{2}}{\hbar^{2}\omega_{q1}^{2}}
4 n_{q1}\sin^2\frac{\omega_{q\kappa}t}{2}
\sim \lim_{q\to 0}
\left\{\frac{|g_{x}^{q1}|^{2}}{\hbar^{2}}n_{q1}\right\}
\frac{L}{v_1}t=\gamma t
\end{equation}
with
\begin{equation} 
\gamma=\frac{k_BT(D_c-D_v)^2(1-2\nu)^2}{2A\rho v_1^3\hbar^{2}}\, .
\end{equation}
Thus, Eq.~(\ref{EqpolarizationExact}) exhibits in the long-time limit an
exponential decay of the polarization $p(t)\sim \e^{-\gamma t}$.
For an isotropic material (as assumed in our numerical calculations), we have 
\[
\nu=\frac{v_L^2-2v_T^2}{2(v_L^2-v_T^2)}
\, ,\quad
\frac{E}{\rho}=\frac{v_T^2(3v_L^2-4v_T^2)}{v_L^2-v_T^2}\, ,
\]
providing a full width at half maximum of $2\hbar\gamma=0.44\mu
\mathrm{eV}\times T/\mathrm{K}$ for the absorption peak in agreement
with the numerical data shown above.
This finite width is in
contrast to the case of three-dimensional phonons 
\cite{KrummheuerPRB2002,ForstnerPSSB2002,ZimmermannICPS2002},
where the $q^2$
factor from the three-dimensional $q$ integral does not allow for
any contribution to the zero-phonon line. 

For our calculations, we considered the phonon spectrum of ideal 
nanowires. For real structures, the presence of the heterostructure
will modify the phonon spectrum, thus possibly changing the location
of the satellite peaks. However, the qualitative nature
of the acoustic branch for
small wavelengths will not be affected. Thus, the broadening mechanism of
the zero-phonon line should be robust, albeit a quantitative change in
the width may occur.

In conclusion, we have calculated the quasi one-dimensional 
phonon spectrum of a
GaAs nanowire and studied its impact on the excitonic 
absorption of an embedded quantum dot.
We find pronounced satellite peaks in the absorption spectrum matching 
extrema in the phonon dispersion, provided that the corresponding 
phonon mode has a non-vanishing radial part. In addition, a finite width of the
zero-phonon line of the order of $0.5\mu\mathrm{eV}\times T/\mathrm{K}$ 
is induced due to the wire elongation mode even for an infinite
phonon lifetime. This is larger than the total width observed in
quantum dots embedded in bulk material \cite{BorriPRB2005} for $T<40$ K.
Thus this effect provides an essential limit for the coherence lifetime of
excitons in nanowire-based quantum dots.

\acknowledgments
This work was supported by the Swedish Research Council (VR) and 
the Deutsche Forschungsgemeinschaft 
(DFG) through Sfb 296.

%\bibliographystyle{apsrev}
%\bibliography{references,help}

\end{document}